**Feedback to the primary visual cortex is highly concentrated on the central visual field representation**

Piotr Majka[1], Emmanuel K L Cho[3], Karolina Łabuszewska[1], Thuy Vy Nguyen[3], Marcin Syc[1],

Tomasz Walkiewicz[1], Katrina H Worthy[3], Li Zhaoping*[2], Marcello Rosa*[3]

[1]Laboratory of Neuroinformatics, Nencki Institute of Experimental Biology of the Polish Academy of Sciences, Warsaw, Poland

[2]University of Tübingen, Max Planck Institute for Biological Cybernetics, Tübingen, Germany

[3]Department of Physiology and Neuroscience Program, Biomedicine Discovery Institute, Monash University, Clayton, Australia

* joint corresponding authors, li.zhaoping@tuebingen.mpg.de marcello.rosa@monash.edu

## Abstract

*Retrograde tracer injections in marmoset monkeys revealed a striking feature of neural circuitry in primate visual system: among neurons projecting to the primary visual cortex (striate cortex, V1), the proportion of extrastriate feedback neurons relative to thalamic feedforward neurons decreases markedly with visual field eccentricity. This supports the Central-Peripheral Dichotomy theory, which proposes fundamentally distinct computational specializations—visual recognition and attentional selection, respectively—in the central and peripheral visual fields.*

The primate visual cortex is formed by areas arranged in a processing anatomical hierarchy[1,2]. This hierarchy begins at V1, where retinal inputs relayed by the lateral geniculate nucleus (LGN) first arrive, and extends a feedforward cascade through V2 and other “downstream” extrastriate areas. These areas, in turn, project feedback connections to V1, whose functional roles remain actively debated[3]. Many anatomical properties, including cell densities, sizes of cortical columns, sizes of neuronal dendritic trees and the density of innervation from the LGN, are approximately constant across the V1 surface, for example between a V1 surface patch representing the central visual field and another representing the peripheral visual fields[4,5,6]. Until recently there has not been a computational motivation to explore the possibility that the feedback neural circuit constitutes an exception.

The Central-Peripheral Dichotomy (CPD) theory[7] proposes that retinotopically organised areas of the cortex are differentiated by a fundamental dichotomy between distinct functions served by the central versus the peripheral visual field, providing a global organizational principle for the brain's visual architecture. This theory was motivated by brain's fundamental processing bottleneck ("attentional bottleneck"), which allows only a tiny fraction of visual inputs to be selected for recognition[8]. According to the CPD theory, computations in the peripheral and central visual fields are specialized, respectively, to select this fraction and to recognize the selected fraction. More specifically, selection leads to gaze or attention shifts to centre the selected visual content at the fovea; subsequently, detailed processing, particularly object recognition, occurs mainly in the central visual field representation.

Assuming that the information bottleneck starts from V1 to downstream areas, the CPD theory additionally proposes that, during ongoing recognition, downstream areas send feedback to query for additional relevant information that can disambiguate between competing perceptual hypotheses. It therefore hypothesizes that the feedback to upstream areas, particularly V1, is concentrated at the representation of the central visual field[7] (Figure 1). This theory has accumulated converging neurophysiological, neural imaging, and behavioural evidence (reviewed in[9]), but there has been no anatomical evidence for its key hypothesis of central-focused feedback.

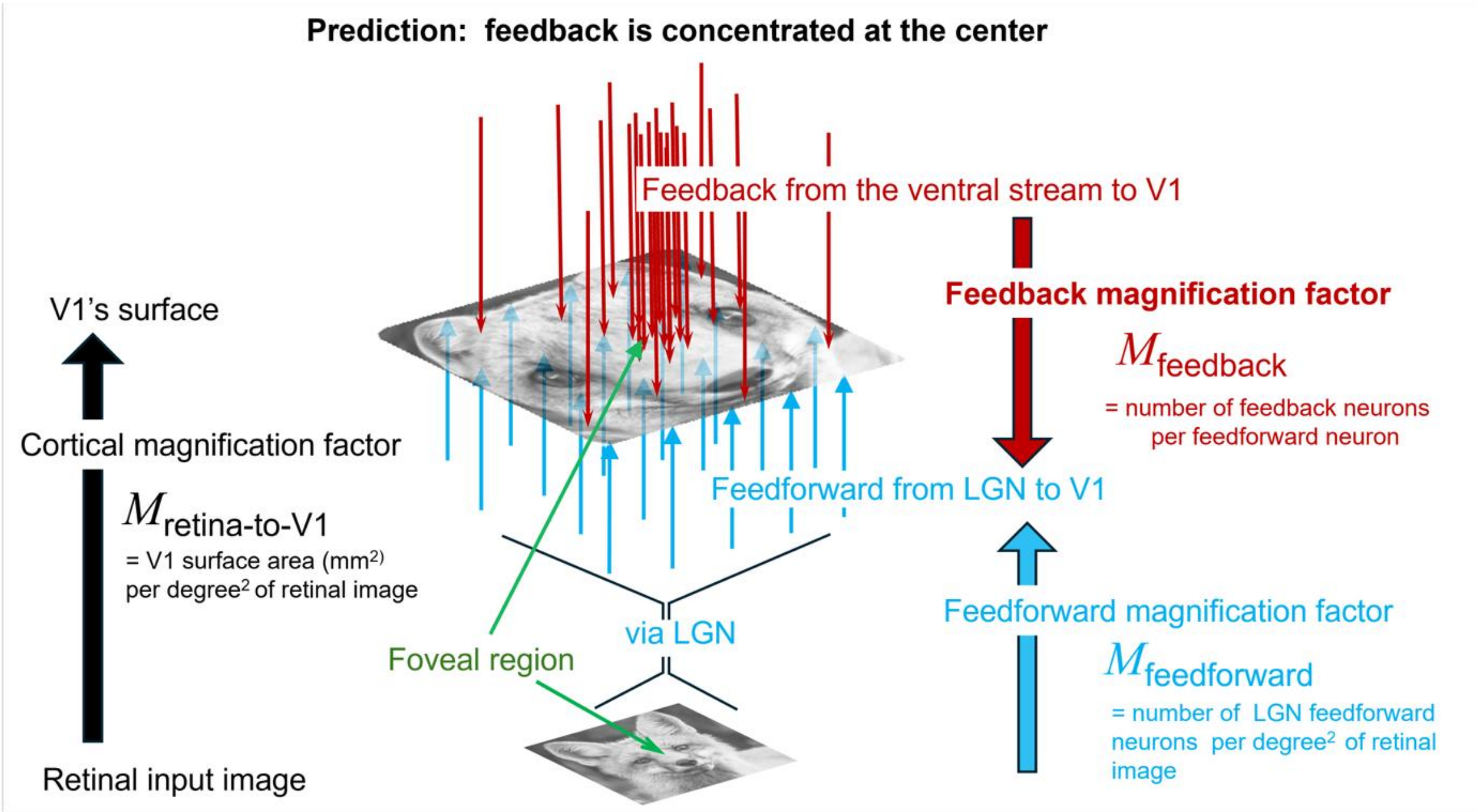

**Figure 1:** A prediction of the Central-Peripheral Dichotomy (CPD) theory: along the visual pathway, top-down feedback to V1 to aid visual recognition is mainly directed to V1's representation of the central visual field. The retinotopic transformation from the retina to V1 (via LGN) expands the central visual field, so that a larger cortical surface area is devoted to each square degree of visual angle in the central than the peripheral visual field[10,11,12]. A cortical magnification factor $M_{\text{retina-to-V1}}$ (in units of $mm^2deg^{-2}$ ) quantifies the amount of V1 surface area devoted to one square degree of visual field at each visual eccentricity $E$. In V1, the density of V1-projecting feedforward or feedback neurons is the number of the neurons in any given brain source region from outside V1 to project to each square millimetre of the V1 surface. The prevalence of the feedback can be quantified by a feedback magnification factor, $M_{\text{feedback}}$ , defined as the ratio of feedback to feedforward projecting neural density. It is a dimensionless value, measuring the number of feedback neurons per feedforward neuron. The CPD theory predicts that this feedback magnification factor $M_{\text{Feedback}}$ in V1 is higher for more central visual field locations. Let $d$ be the density of LGN innervation of V1, defined as the number of LGN neurons projecting to each square millimeter of V1 surface. Then, a feedforward magnification factor, $M_{\text{feedforward}} \equiv M_{\text{retina-to-V1}} \times d$, gives the number of feedforward LGN neurons to V1 per square degree of visual angle. The number of feedback neurons to V1 surface devoted to one square degree of visual field at eccentricity $E$ is then $N(E)=M_{\text{feedback}}(E) \cdot M_{\text{feedforward}}(E) \equiv M_{\text{feedback}}(E) \cdot M_{\text{retina-to-V1}}(E) \cdot d(E)$. Let $E_1$ and $E_2$ be the eccentricities of a central and a peripheral location, respectively. If $\mathrm{M_{feedback}(E_1)/M_{feedback}(E_2)}$=10 and $\mathrm{M_{retina\text{-}to\text{-}V1}(E_1)/M_{retina\text{-}to\text{-}V1}(E_2)}$=100, then, $\mathrm{N(E_1)/N(E_2)}$=1000, or more precisely, $1000 \times \mathrm{d(E_1)/d(E_2)}$ (typically $\mathrm{d(E_1)} \approx \mathrm{d(E_2)}$ since LGN innervation density to V1 surface is approximately uniform[5,13]. The number of feedback neurons projecting to $1\ \mathrm{mm}^2$ of V1 surface patch is $\mathrm{M_{feedback}(E)} \cdot \mathrm{d(E)}$ and is predicted to be larger at smaller eccentricities $\mathrm{E}$. For conceptual clarity, this schematic merges the two V1 hemispheres representing the left and right visual fields.

Here we tested the critical prediction that the feedback to V1 is focused on the central field representation. To this end, we injected cellular-resolution retrograde tracers at twelve V1 sites of six marmoset monkeys, which spanned the representations of visual field eccentricity $E$ from $2^o$ to $18^o$. The tracer was taken up by presynaptic axon terminals and retrogradely transported to label the cell bodies of neurons that project monosynaptically to each injection site. Because the absolute numbers of neurons labelled by tracer injections vary according to factors such as the effective spread and chemical properties of the tracer, the quantitative numbers should be normalised in a manner that allows valid comparisons across injection sites and animals. Therefore, for each source brain region (e.g., cortical areas such as V2, V4 and MT, or lateral geniculate nucleus (LGN)), we first quantified the number $N_{\text{brain region}}(E)$ of labelled neurons arising from a V1 injection at eccentricity $E$[14]. Then, we defined a normalized measure,

$$\hat{N}_{\text{brain region}}(E) \equiv \frac{N_{\text{brain region}}(E)}{N_{\text{LGN}}(E)},$$

which is the number of feedback neurons projecting from this brain region per feedforward neuron projecting from the LGN. Note that both $N_{\text{brain region}}(E)$ and $N_{\text{LGN}}(E)$ scale with the spread and uptake of the injected tracer. Hence, our measure $\hat{N}_{\text{brain region}}(E)$ normalizes away variations in the tracer spread and uptake across our injections, providing a robust index to quantify the feedback magnification factor $M_{\text{Feedback}}$ for a given feedback source brain region to V1. As illustrated in the Figures, $M_{\text{feedback}}$ quantifies the brain's further magnification to a visual field location via the feedback, beyond the feedforward magnification via the cortical magnification factor $M_{\text{retina-to-V1}}$ (or via the feedforward magnification factor $M_{\text{feedforward}}$).

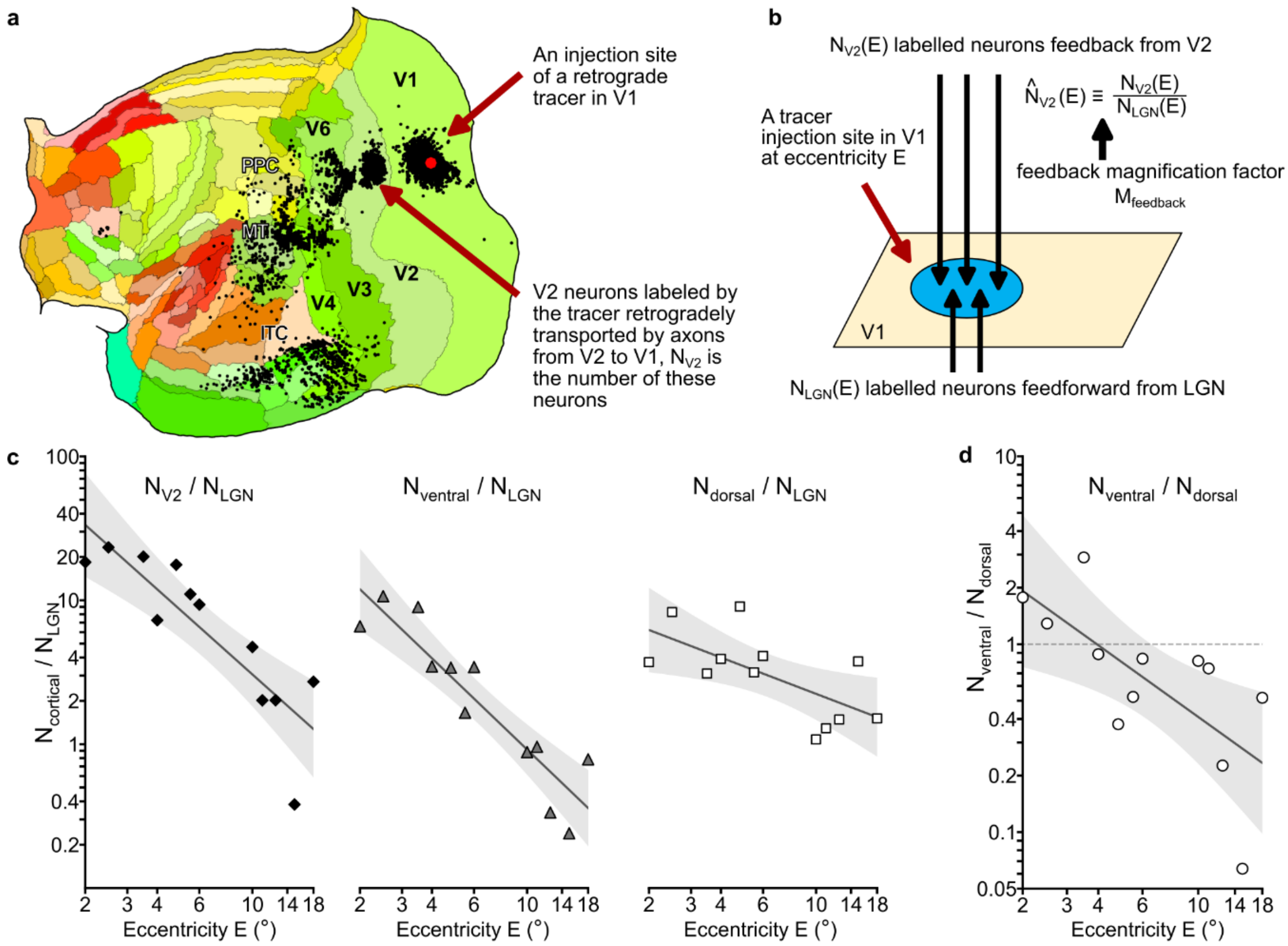

**Figure 2:** Methods and results of the anatomical test of the prediction from the central-peripheral dichotomy (CPD) theory: feedback to V1 to aid recognition is mainly directed to the representation of the central visual field. **a**: tracer-labelled neurons in various cortical areas as the result of an injection of a retrograde tracer to a V1 site, illustrated in an unfolded representation of the cortex[15]. Each black dot marks a neuron which projects axon(s) to the injected site (red circle) in V1. The number of such dots in each area outside V1 was obtained, e.g., $N_{\text{V2}}$ is the number of labelled V2 neurons feeding back to the injection site in V1. This map and corresponding materials from other injections, can be found at www.marmosetbrain.org[15]. **b**: for each injection site, we also obtained the number $N_{\text{LGN}}$ of tracer-labelled neurons in lateral geniculate nucleus (LGN), the source of feedforward inputs to V1. For each cortical region, e.g., V2, the ratio $\hat{N}_{\text{V2}}(E) \equiv N_{\text{V2}}/N_{\text{LGN}}$, or normalized number,

is obtained at an injection site of eccentricity $E$ to measure the feedback magnification factor $M_{\text{feedback.}}$ **c**: $M_{\text{feedback}}$ for V2, ventral stream areas, and dorsal stream areas. Across twelve injection sites with eccentricities $2^o \leq E \leq 18^o$, the normalized number $\hat{N}_{\text{V2}}(E)$ of feedback neurons in V2 decreased with $E$ by an order of magnitude. A similar trend is seen in $\hat{N}_{\text{ventral}}$, the normalized number of tracer-labelled neurons pooled across visual cortical areas in the ventral stream. The decreasing trend is weaker in $\hat{N}_{\text{dorsal}}$, the normalized number of labelled neurons pooled across visual cortical areas in the dorsal stream. The three solid lines are linear fits of $log\hat{N}_{\text{brain region}}$ versus $logE$ in the brain regions V2, ventral stream, and dorsal stream. The slope (and its 95% confidence interval) of each line is $\alpha$=1.5 (0.86, 2.1) for V2, $\alpha$=1.6 (1.1, 2.1) for the ventral stream, and $\alpha$=0.6 (0.13, 1.15)for the dorsal stream, giving a power law model $\hat{N}_{\text{brain region}}(E)$=$constant \cdot E^{-\alpha}$. The shading around each line indicates the 95% confidence interval of the fit. **d**: like **c**, now for $N_{\text{ventral}}/N_{\text{dorsal}}$ (or equivalently $\hat{N}_{\text{ventral}}/\hat{N}_{\text{dorsal}}$) versus $E$, with $\alpha$=0.96 (0.26, 1.7).

We found that, for feedback from area V2, the normalized number $\hat{N}_{V2}(E)$ decreased with $E$ approximately by a power law

$$\hat{N}_{V2}(E) \approx constant \cdot E^{-\alpha}, \alpha=1.5 \in \text{ a 95\% confidence interval } (0.86, 2.1),$$

reducing $\hat{N}_{V2}(E)$ by roughly 10 times over our eccentricity range (Figure 2**c**).

Since the CPD theory posits that the feedback supports visual recognition by disambiguating between competing perceptual hypotheses, and since recognition is arguably more linked to the ventral stream of the visual pathway[16], we also pooled the labelled neurons across all the visual cortical areas (excluding V2) in this stream to obtain $N_{\text{ventral}}(E)$. Like in V2, the normalized number

$$\hat{N}_{\text{ventral}}(E) \approx constant \cdot E^{-\alpha}, \alpha=1.6 \in \text{ a 95\% confidence interval } (1.1, 2.1),$$

decreases also by about 10 times within the eccentricity range. These findings confirm our main prediction from the CPD theory.

Neurons in the dorsal stream also send feedback to V1. Because object recognition is arguably a role more aligned with the ventral stream, we explored the possibility that the feedback from dorsal stream areas to V1 would follow a quantitatively different pattern. Indeed, the normalized number of labelled neurons pooled across dorsal stream areas $\hat{N}_{\text{dorsal}}(E) \propto E^{-\alpha}$ decreases less steeply with increasing $E$ (Figure 2**c**), by a smaller $\alpha$=0.6 in a 95% confidence interval $(0.13, 1.15)$. Figure 2**d** shows

$$\frac{N_{\text{ventral}}(E)}{N_{\text{dorsal}}(E)} \approx constant \cdot E^{-\alpha}, \alpha=0.96 \in \text{ a 95\% confidence interval } (0.26, 1.7),$$

so that, excluding V2, the dominant feedback source to V1's representation of the central and peripheral field is the ventral and dorsal stream, respectively.

Since the CPD theory posits that the central and peripheral visual fields are specialized for visual recognition and attentional selection (mainly by gaze shifts), respectively, our data is consistent with the idea that the ventral stream is more specialized for "what" (recognition) processing whereas the dorsal stream is more specialized for "where" objects are[16], a process useful for gaze shifts. In light of the CPD theory, the above observations suggest that the dorsal stream may also engage significantly in recognition, although not as much as the ventral stream does, likely extracting properties such as size, orientation, and motion direction of objects to guide actions[17].

Finally, V1 also receives several quantitatively less numerous feedback projections from higher-level areas that are not theoretically aligned with either the dorsal or ventral streams, including areas in the parahippocampal, auditory and posterior cingulate/ retrosplenial regions[18,19]. For comparison, the normalized number $\hat{N}_{\text{non-visual cortical regions}}(E)$ for non-visual cortical regions pooled together does not consistently increase or decrease with $E$, giving , $\alpha \approx 0$ (Supplementary Figure 1).

There has been previous indirect evidence in support of the critical CPD prediction of centrally focused feedback. This includes a central bias in the functional connectivity between frontal areas and V1[20] and a foveal focus of inferred feedback from V4 to V1 (using Granger causality analysis of neural activities[21]). Other support comes from psychophysical studies of the reversed depth illusion and the flip tilt illusion, which are typically visible only in the peripheral visual field, as predicted by the CPD theory[9,22]. These illusions have been interpreted as perceptual errors arising from a lack of the feedback query to aid recognition in light of the bottleneck. As predicted, they become visible in the central visual field when the feedback query is disrupted by backward masking.

In addition to their relevance for understanding the interplay between central and peripheral vision during primate behaviour, our anatomical results challenge the long-standing view that each retinotopic visual area can be conceptualized as an approximately uniform surface in computational terms. In this view, uniformly defined compartments within a given retinotopic cortical area (e.g., a set of columns within 1 $mm^2$ of cortical surface) would perform the same kind of processing, differing only in cortical magnification factors (e.g., $M_{\text{retina-to-V1}}$ for V1) that scale inversely with the sizes of the visual field regions being processed[4,23]. This has been traditionally supported by observations that visual behavioural performances or neural receptive properties at different eccentricities can be equated with each other by scaling the sizes of visual inputs or receptive fields[24].

However, this long-standing view has been questioned by other data, including task-dependence in the scaling of visual input sizes with eccentricity to equate behavioural performances[24]. There is a growing recognition of an eccentricity bias, or central-peripheral dichotomy, in behavioural data (e.g. in visual crowding[25] and perceptual organization[26]), particularly when higher cortical areas and more complex visual tasks are concerned[27,28].

Motivated by the CPD theory, our retrograde tracing data provides a new perspective, revealing an order-of-magnitude change in the feedback prevalence, quantified by the feedback magnification factor $M_{\text{feedback}}$, across eccentricity, and thereby supporting the fundamentally different roles for the central and peripheral visual fields. Even in V1, arguably the best-defined visual area, eccentricity-dependent subregions specialize to serve distinct functions. The CPD theory offers a computational foundation to link neural with behavioural data. It motivates a framework to view our brain's visual architecture from two different axes: a traditional one along the visual hierarchy and a new axis along the central-peripheral dichotomy.

**Acknowledgements**

The authors would like to acknowledge the contributions of Shi Bai, who is behind the computational infrastructure serving the Marmoset Brain Architecture Project. Alice Nguyen, Alice Cavazzuti and Vincent Day contributed to the microscopy work underlying the collection of data. We also thank Cecilia Cranfield for administrative work in support of this project.

Grant support:

National Science Centre Poland (2019/35/D/NZ4/03031, 2025/56/Q/NZ2/00227)

National Health and Medical Research Council (1194206, 2041019)

Australian Research Council (DP210101042, DP210103865)

Funding from the University of Tübingen and the Max Planck Society.

# Online Methods

**Animals:** All experiments adhered to the Australian Code of Practice for the Care and Use of Animals for Scientific Purposes and strictly followed protocols approved by the Monash University Animal Experimentation Ethics Committee. To minimize the number of animals used as part of ongoing and previous experiments each animal also received tracer injections in other brain regions (to be reported separately). The results derive from 12 retrograde tracer injections placed in the striate cortex of 6 young adult marmoset monkeys (Supplementary Table S1; for full datasets including detailed metadata about each animal see www.marmosetbrain.org). Following histological reconstruction and computerized reconstruction (see below), the locations of these injection sites were determined relative to areas according to the Paxinos atlas[29].

**Surgical procedures and retrograde tracer injections:** The surgical protocols have been detailed in previous papers[15]. The animals were fasted the night before and oral antibiotics (Cephalexin 30 mg/kg) and analgesics (meloxicam 0.1mg/kg) were also administered. They were pre-medicated with intramuscular (i.m.) injections of atropine (0.2 mg/kg) and diazepam (2 mg/kg), and then, 20-30 minutes later, anesthetised with alfaxalone (8 mg/kg, i.m.). Under anaesthesia they also received injections of dexamethasone (0.3 mg/kg, i.m.) and were fitted to a stereotaxic frame. The eyes were protected by topical application of GenTeal gel (hypromellose 0.3%). Vital signs such as body temperature, heart rate, and partial blood oxygenation were monitored. To maintain areflexic state additional anaesthetic doses of alfaxalone were administered as necessary.

Following a craniotomy the dura mater was incised. The tracers were injected using a microsyringe, guided by stereotaxic coordinates[10]. After the tracers were injected, the dural flaps were repositioned over the cortical surface, and then covered with moistened ophthalmic film. Then, the excised bone fragment was repositioned and sealed using dental acrylic. Post-operative analgesics were administered immediately after the animal regained consciousness or displayed spontaneous movements (Buprenorphine 0.01mg/kg, s.c.). Analgesics were continued for the following 2–3 days (Meloxicam 0.05mg/kg, oral) and antibiotics for 5 days (Cephalexin 30 mg/kg, oral).

**Histological processing:** Following survival times (Table 1), the animals were re-anesthetised with alfaxalone (12mg/kg, i.m.) and, after attaining full areflexia, received a lethal dose of pentobarbitone sodium (100 mg/kg, i.p.). Upon respiratory arrest, they were perfused through the heart with heparinised saline (1L), followed by 4% paraformaldehyde in 0.1 M phosphate-buffered saline (pH 7.4, 1L). The brains were removed from the cranium and postfixed for a minimum of 24 hours before being cryoprotected by immersion in increasing concentrations of sucrose (10 to 30%, in 4% buffered paraformaldehyde). They were then sectioned at 40-µm in the coronal plane, using a cryostat. Five series of sections were obtained. To examine the fluorescent labelling all sections from one series were mounted unstained, and the other four series of sections were stained for Nissl substance, myelin and cytochrome oxidase (CO) activity. After all sections were air dried, they were cover slipped with di-n-butyl phthalate xylene (DPX).

Estimates of the eccentricity of injection sites were done by reference to stereotaxic coordinates resulting from the computational reconstruction process (see below), relative to receptive fields illustrated in serial sections[10].

**3D Reconstruction and Co-registration to the Nencki–Monash Template Space:** Stained sections were scanned using an Aperio Scanscope AT Turbo system (Leica Biosystems) at a resolution of 0.50 µm per pixel. To register individual experimental datasets into a common reference space, serial coronal Nissl-stained section images (149–154 sections per case) were downsampled to an isotropic 40 µm per pixel resolution. They were then pre-processed, and reconstructed into three-dimensional (3D) volumes using previously established computational protocols[14,15,30]). Reconstruction was performed using a pipeline based on the open-source Possum framework[31] and Advanced Normalization Tools (ANTs; Avants et al., 2011; http://stnava.github.io/ANTs/; RRID:SCR_004757).

Recovery of a global 3D anatomical shape was achieved with an iterative affine pipeline that alternated between 3D affine registration of the reference volume and 2D rigid section alignment driven by Mattes Mutual Information. Section-specific non-linear distortions (e.g., local tissue stretching or compression) were subsequently corrected with a 2D deformable reconstruction algorithm.

Individual 3D hemispheric reconstructions were spatially co-registered into the standardized stereotaxic space of the Nencki–Monash template (NM template, RRID:SCR_018367)[32]. This was performed using a combination of global 3D affine alignment followed by non-linear 3D deformation[32]. Reconstructed volumes and template images were resampled to 75 µm isotropic resolution. To preserve anatomical fidelity and mitigate inter-individual cortical variation, spatial registration was guided by two equally weighted image similarity metrics: cross-correlation (CC[33]) for Nissl optical density images and Point-Set Expectation (PSE[34]) for structural and cytoarchitectural label maps (e.g., areas V1, MT, A3b, Ent, Pir) that were manually outlined across histological series by an expert neuroanatomist (21 to 27 areas)[15,30,32].

Retrogradely labelled cell bodies and injection site boundaries were plotted from adjacent fluorescent section series under epifluorescence microscopy and aligned with corresponding Nissl sections[14]. The computed 2D and 3D spatial transformation fields established a continuous, bidirectional mapping between raw histological coordinates and the stereotaxic space of the NM template. Using these spatial displacement fields, individual 3D cell coordinates were mapped into template space and automatically assigned to specific cortical areas based on the 116-area cytoarchitectural parcellation scheme[29] incorporated into the NM template[32]. Summing the mapped neurons assigned to each area yielded absolute cell counts for each cortical area.

**Statistical analysis:** Since, according to the CPD theory, a key role of feedback connections is to query for information that helps disambiguate between competing perceptual hypotheses, it was important to distinguish feedback connections from different cortical areas according to their roles in object recognition. For this purpose, we classified all cortical areas projecting to V1 according to whether they are classically seen as aligned with the ventral stream or the dorsal stream (Ungerleider and Mishkin; Goodale and Milner 1992). Supplementary Table S2 summarises all sources of feedback connections to V1 observed across all 12 injection sites, and their classification labels as ventral stream, dorsal stream, or other. Connections from V2 were analysed separately, since V2 is not only numerically the most prominent source of feedback to V1 (Figure 2), but is also related to both streams[35]. Other cortical areas, for which a role more aligned with the classical streams has not been determined, were analysed separately.

In Figure 2**c**, the three solid lines are the results of fitting the linear function

$$log\hat{N}_{\text{brain region}}(E) = -\alpha\ logE + \text{intersect}$$

to obtain the slope $\alpha$ and its 95% confidence interval using matlab function fit. This fit gives the power law $\hat{N}_{brain\ region}(E) \propto E^{-\alpha}$. One-sided Student's $t$-test was used to obtain the probability, i.e., the *p*-value, that the positive $\alpha$ (the result of the fitting) comes from the null hypothesis of $\alpha$=0. The *p*-values were $p$=0.000, 0.000, and 0.01 for data from V2, ventral stream, and dorsal stream, respectively. The same procedure was used to fit data in Figure 2**d** for the power law $\frac{N_{\text{ventral}}(E)}{N_{\text{dorsal}}(E)} \propto E^{-\alpha}$, giving *p*-value $p$=0.006 that the positive $\alpha$ could come from the null hypothesis of $\alpha$=0.

**Additional references for online methods:**

# Supplementary information

**Supplementary figure 1**

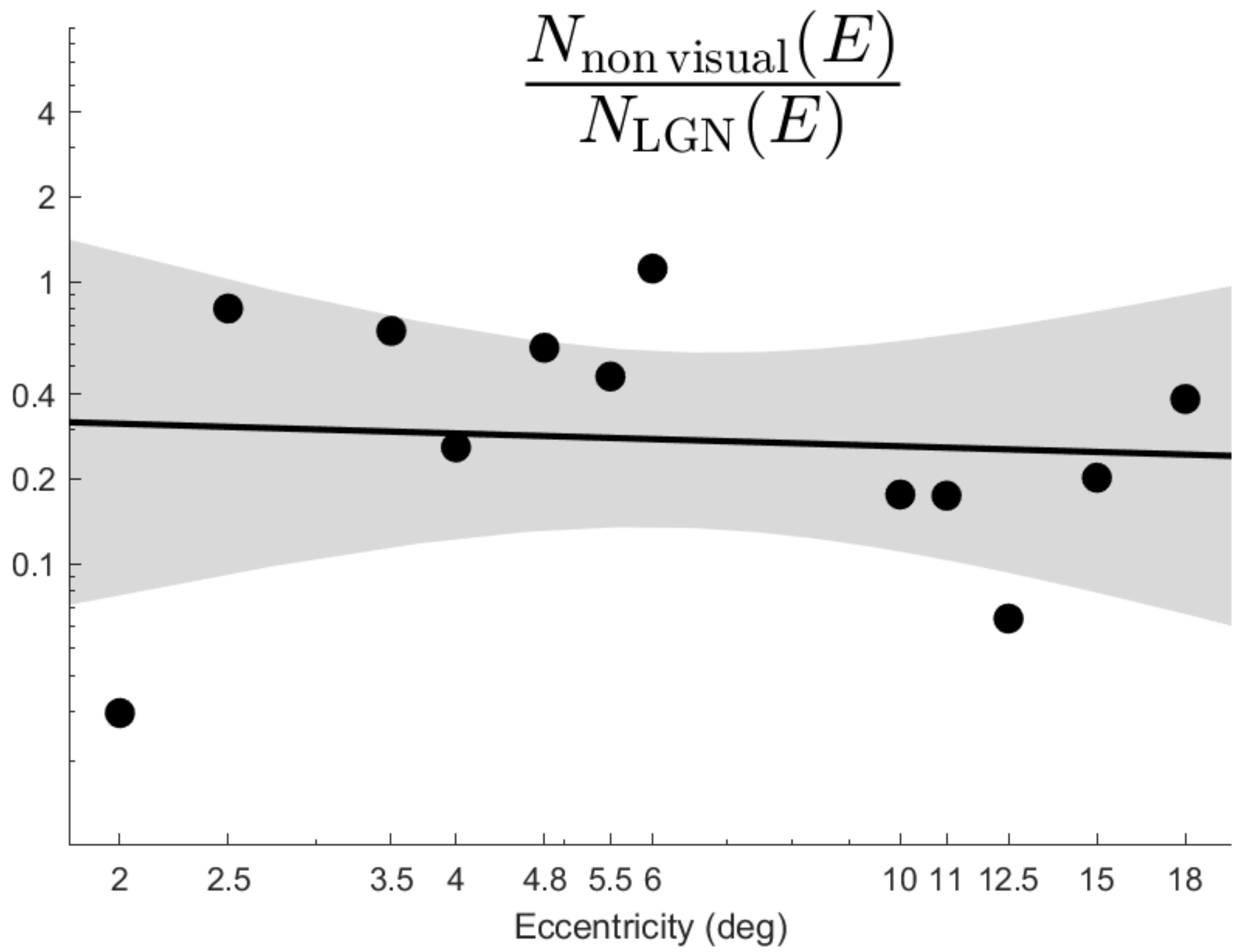

**Supplementary Table 1: Animals and injection sites**

| Animal | Tracer[a] | Sex | Age | Survival (days) | Hemisphere | Eccentricity | Stereotaxic coordinates AP (A) (mm) | ML (L) (mm) | DV (H) (mm) |
|---|---|---|---|---|---|---|---|---|---|
| CJ240 | DY | M | 2 years, 11 months | 13 | R | 2.0˚ | -9.3 | +5.9 | +9.2 |
| CJ240 | CTBr | M | 2 years, 11 months | 13 | R | 2.5˚ | -9.9 | +4.9 | +11.4 |
| CJ243 | FB | M | 2 years, 8 months | 17 | R | 3.5˚ | -11.2 | +0.5 | +11.5 |
| CJ240 | FB | M | 2 years, 11 months | 13 | R | 4.0˚ | -11.1 | +1.2 | +9.8 |
| CJ240 | CTBg | M | 2 years, 11 months | 13 | R | 4.8˚ | -8.9 | +0.7 | +14.3 |
| CJ82 | FE | F | 2 years, 4 months | 15 | L | 5.5˚ | -8.6 | +2.3 | +5.2 |
| CJ243 | CTBr | M | 2 years, 8 months | 17 | R | 6.0˚ | -9.9 | +1.1 | +3.7 |
| CJ82 | FR | F | 2 years, 4 months | 15 | L | 10.5˚ | -5.6 | +0.6 | +13.0 |
| CJ174 | FB | F | 2 years, 6 months | 16 | R | 12.0˚ | -8.8 | +3.0 | +11.3 |
| CJ243 | DY | M | 2 years, 8 months | 17 | R | 13.0˚ | -8.2 | +4.2 | +12.1 |
| CJ178 | FB | F | 2 years, 2 months | 18 | R | 15.0˚ | -7.5 | +3.5 | +13.0 |
| CJ238 | FB | F | 2 years, 5 months | 9 | R | 18.0˚ | -5.5 | +5.0 | +12.5 |

**Supplementary Table 2: Classification of areas projecting to V1 within streams**

| Cortical sector (Paxinos et al. 2012) | Area name | Abbreviation | Classification | Total labelled cells (12 injections) |
|---|---|---|---|---|
| Dorsolateral prefrontal cortex | Area 46, ventral part | A46V | other | 1 |
| | Area 8a, dorsal part | A8aD | other | 2 |
| | Area 8a, ventral part | A8aV | other | 31 |
| Ventrolateral prefrontal cortex | Area 47 (12), lateral part | A47L | other | 3 |
| | Area 45 | A45 | other | 3 |
| | Proisocortical motor region (precentral opercular area) | ProM | other | 1 |
| Orbitofrontal cortex | Area 13, medial part | A13M | other | 1 |
| Motor and premotor cortex | Area 8, caudal part | A8C | other | 1 |
| Insula and rostral lateral sulcus cortex | Agranular insular area | AI | other | 1 |
| | Dysgranular insular area | DI | other | 2 |
| | Insular proisocortex | IPro | other | 3 |
| | Temporal proisocortex | TPro | other | 3 |
| Somatosensory cortex | Secondary somatosensory area, parietal ventral area | S2PV | other | 2 |
| | Secondary somatosensory area, internal part | S2I | other | 7 |
| | Secondary somatosensory area, external part | S2E | other | 9 |
| Auditory cortex | Rostral auditory area | R | other | 1 |
| | Primary auditory area | A1 | other | 49 |
| | Caudomedial area | CM | other | 86 |
| | Caudal parabelt area | CPB | other | 16 |
| | Middle lateral area | ML | other | 29 |
| | Caudolateral area | CL | other | 37 |
| Lateral and inferior temporal cortex | Temporal area TE, part 1 | TE1 | ventral stream | 2 |
| | Temporo-parieto-occipital association area | TPO | other | 97 |
| | Retroinsular area | ReI | other | 3 |
| | Temporal area TE, part 2 | TE2 | ventral stream | 39 |
| | Area PGa andIPa (fundus of superior temporal ventral area) | Pga/IPa | other | 84 |
| | Temporoparietal transitional area | TPt | other | 51 |
| | Temporal area TE, part 3 | TE3 | ventral stream | 98 |
| | Temporal area TE, occipital transition part | TEO | ventral stream | 557 |
| Ventral temporal cortex | Entorhinal area | Ent | other | 1 |
| | Area 36 | A36 | other | 22 |
| | Temporal area TF | TF | other | 338 |
| | Temporal area TL | TL | other | 146 |
| | Temporal area TH | TH | other | 501 |
| | Temporal area TL, occipital transition part | TLO | ventral stream | 884 |
| | Temporal area TF, occipital transition part | TFO | ventral stream | 1075 |
| Posterior cingulate, medial, and retrosplenial cortex | Area 23a | A23a | other | 8 |
| | Area 30 | A30 | other | 5 |
| | Area 23b | A23b | other | 3 |
| | Area 29, parts a-c | A29a-c | other | 9 |
| | Area 23, ventral part | A23V | other | 678 |
| | Prostriate area | ProSt | other | 442 |
| Posterior parietal cortex | Parietal area PF | PF | dorsal stream | 5 |
| | Parietal area PE | PE | dorsal stream | 18 |
| | Parietal area PFG | PFG | dorsal stream | 59 |
| | Area 31 | A31 | other | 27 |
| | Anterior intraparietal area | AIP | dorsal stream | 29 |
| | Parietal area PG | PG | dorsal stream | 218 |
| | Parietal area PE, caudal transitional part | PEC | dorsal stream | 27 |
| | Ventral intraparietal area | VIP | dorsal stream | 20 |
| | Lateral intraparietal area | LIP | dorsal stream | 481 |
| | Parietal area PG, medial part | PGM | dorsal stream | 52 |
| | Visual area 6a (posterior parietal medial area) | V6a | dorsal stream | 154 |
| | Occipito-parietal transitional area | OPt | dorsal stream | 672 |
| | Medial intraparietal area | MIP | dorsal stream | 117 |
| | | | | |
| Visual cortex | Medial superior temporal area | MST | dorsal stream | 1236 |
| | Fundus of superior temporal sulcus area | FST | dorsal stream | 669 |
| | Visual area 5 (middle temporal area, MT) | V5 | dorsal stream | 5674 |
| | Visual area 4, transitional part (middle temporal crescent) | V4T | dorsal stream | 1832 |
| | Area 19, medial part | A19M | dorsal stream | 4209 |
| | Visual area 3a (dorsoanterior area, DA) | V3a | dorsal stream | 427 |

| | | | |
|---|---|---|---|
| Visual area 4 (ventrolatereral anterior area, VLA) | V4 | ventral stream | 4802 |
| Visual area 6 (dorsomedial area, DM) | V6 | dorsal stream | 10849 |
| Area 19, dorsointermediate part (DI) | A19DI | dorsal stream | 1466 |
| Visual area 3 (ventrolateral posterior area, VLP) | V3 | ventral stream | 10097 |
| Visual area 2 | V2 | V2 | 49840 |